\documentclass[1p,sort&compress,times]{elsarticle}
\usepackage{amsmath,amssymb,amscd,amsthm}
\makeatletter
\def\ps@pprintTitle{%
 \let\@oddhead\@empty
 \let\@evenhead\@empty
 \def\@oddfoot{}%
 \let\@evenfoot\@oddfoot}
\makeatother
\usepackage[utf8]{inputenc}
\usepackage[T1]{fontenc}
\usepackage{lmodern}
\usepackage[colorlinks,linkcolor=blue,citecolor=blue,urlcolor=blue]{hyperref}
\newcommand{\al}{\alpha}
\newcommand{\be}{\beta}

\newcommand{\ga}{\gamma}

\newcommand{\la}{\lambda}
\newcommand{\om}{\omega}
\newcommand{\si}{\sigma}

\newcommand{\vp}{\varphi}

\newcommand{\ze}{\zeta}
\newcommand{\bx}{\mathbf{x}}

\newcommand{\RR}{{\mathbb R}}
\newcommand{\ZZ}{{\mathbb Z}}
\newcommand{\mss}{\kern 1pt}

\renewcommand{\le}{\leqslant}

\newcommand{\tends}[1]{\bbuildrel{\hbox to 2em{\rightarrowfill}}_{#1}^{}}

\newcommand{\iu}{\mathrm i}
\newcommand{\e}{\mathrm e}
\newcommand{\diff}{\mathrm{d}}
\newcommand{\dd}{\mathrm{d}}
\newcommand{\Or}{\mathrm O}
\newcommand{\qbinom}[3]{{#1\atopwithdelims[]#2}_{\raise 3pt\hbox{$\scriptstyle #3$}}}
\renewcommand{\Im}{\operatorname{Im}}

\newcommand{\pdf}[2]{\frac{\partial #1}{\partial #2}}

\newcommand{\Int}[1]{\,\mathop{\!#1}\limits^{\lower1ex\hbox{$\scriptstyle\circ$}}{}}

\theoremstyle{remark}

\newcommand{\mathclap}[1]{\hbox to0pt{\hss$\scriptstyle #1$\hss}}

\def\clap#1{\hbox to 0pt{\hss#1\hss}}

\def\mathclap{\mathpalette\mathclapinternal}

\def\mathclapinternal#1#2{%
  \clap{$\mathsurround=0pt#1{#2}$}}

\begin{document}
\begin{frontmatter}
\title{Jastrow-like ground states for quantum many-body potentials with near-neighbors
  interactions}

\author{Marzieh Baradaran} \ead{marzie.baradaran@yahoo.com}
\address{Department of Physics, University of Guilan, Rasht 41635-1914, Iran}
\author{Jos\'e A.~Carrasco} \ead{joseacar@ucm.es}
\author{Federico Finkel} \ead{ffinkel@ucm.es}
\author{Artemio Gonz\'alez-L\'opez\corref{cor}} \ead{artemio@ucm.es}

\cortext[cor]{Corresponding author}

\address{Departamento
  de F\'{\i}sica Te\'{o}rica II, Universidad Complutense de Madrid, 28040~Madrid, Spain}

\date{September 29, 2017}

\begin{abstract}
  We completely solve the problem of classifying all one-dimensional quantum potentials with
  nearest- and next-to-nearest-neighbors interactions whose ground state is Jastrow-like, i.e., of
  Jastrow type but depending only on differences of consecutive particles. In particular, we show
  that these models must necessarily contain a three-body interaction term, as was the case with
  all previously known examples. We discuss several particular instances of the general solution,
  including a new hyperbolic potential and a model with elliptic interactions which reduces to the
  known rational and trigonometric ones in appropriate limits.
\end{abstract}

\end{frontmatter}
\numberwithin{equation}{section}

\section{Introduction}

Since their introduction in the early 70's, the quantum integrable many-body models of
Calogero~\cite{Ca71} and Sutherland~\cite{Su71,Su72} have been extensively studied due to their
conceptual simplicity and their outstanding properties. In fact, the fundamental character of
these models is attested by their appearance in such diverse areas as soliton
theory~\cite{Po95,AGK11}, orthogonal polynomials~\cite{BF97,Du98,Ta05}, random matrix
theory~\cite{TSA95,St04,CM04}, fractional statistics and anyons~\cite{CL99,Po06}, quantum Hall
effect~\cite{AI94,BH08}, conformal field theory~\cite{Ca04,SSAFR05,EPSS12}, general
relativity~\cite{GT99,Ga12}, hydrodynamics of cold atomic gases~\cite{KA12}, and quantum
quenching~\cite{RS14}.

A remarkable feature of the Calogero and Sutherland models is that their ground-state
wave function~$\psi$ is factorized over the $A_{N-1}$ root system, i.e., is of the form
\begin{equation}\label{CSfactor}
  \psi(\bx)\propto\prod_{i=1}^N\rho(x_i)\cdot\prod_{\mathclap{1\le i<j\le N}}\chi(x_i-x_j).
\end{equation}
Here~$\bx\equiv(x_1,\dots,x_N)$ and
\[
  \rho(x)=\e^{-\frac12\om x^2},\qquad \chi(x)=|x|^a
\]
for the (harmonic) Calogero model, while
\[
  \rho(x)=1\,,\qquad \chi(x)=|\sin x|^a
\]
for the Sutherland model (with $\om>0$ and~$a>-1/2)$. As first noted by
Sutherland~\cite{Su71,Su71b}, this property makes it possible to compute in closed form certain
correlation functions of the latter models by exploiting their connection with random matrix
theory. Indeed, for Calogero's model~$\psi^2$ coincides with the joint probability density of the
eigenvalues of the Gaussian orthogonal, unitary and symplectic ensembles respectively for
$2\om=2a=1,2,4$, while the same relation holds for the ground state of the Sutherland model and
Dyson's circular unitary ensembles (with eigenvalues parametrized as $\e^{2\iu x_k}$)~\cite{Dy62}.
Later on, Dyson~\cite{Dy72} showed how to construct analogues of the Gaussian ensembles with
eigenvalues distributed according to~$\psi^2$ in Eq.~\eqref{CSfactor}, with essentially
arbitrary~$\rho$ and~$\chi(x)=|x|^{\be/2}$ (as usual, $\be$ will be assumed to take the
values~$1,2,4$ for the orthogonal, unitary and symplectic ensembles, respectively).

In view of these results, it is natural to look for the most general quantum Hamiltonian of
Calogero--Sutherland (CS) type (i.e, with one- and two-body long-range interactions) whose ground
state is of the form~\eqref{CSfactor}. A restricted version of this problem (with~$\rho=1$) was
already formulated by Sutherland himself~\cite{Su71}, who later found a solution thereof with an
elliptic two-body interaction potential~\cite{Su75prl,Su75bprl}. Shortly afterwards,
Calogero~\cite{Ca75} showed that this is in fact the most general solution of this restricted
problem. The general problem (with~$\rho$ not necessarily equal to~$1$) was tackled by Inozemtsev
and Meshcheryakov~\cite{IM84}, who claimed to have found a complete solution. A decade later,
however, Forrester~\cite{Fo94jsp} found a model of CS type whose ground state, which exhibits
long-range crystalline order in the thermodynamic limit, is of the factorized
form~\eqref{CSfactor} and yet did not appear in the classification of Ref.~\cite{IM84}. The latter
classification was finally completed several years later by Koprucki and Wagner~\cite{KW00}, who
obtained Forrester's model as a particular case.

The probability distribution $p_\be(s)$ of the (normalized) spacing $s$ between two consecutive
eigenvalues of the Gaussian $\be$-ensembles is approximately given by Wigner's
surmise~$p_\be(s)=A_\be s^\be\e^{-c_\be s^2}$, where the positive parameters~$A_\be,c_\be$ are
fixed by normalization and the condition that the mean spacing be equal to $1$ (see, e.g.,
Refs.~\cite{Me04,Fo10}).
By contrast, it has been conjectured~\cite{BT77} that the spacings distribution of a ``generic''
quantum integrable model is Poissonian, i.e., $p(s)=\e^{-s}$. The latter distributions are
actually obeyed by the spectra of a wide range of either fully chaotic or completely integrable
systems~\cite{Me04}. However, for certain so-called pseudo-integrable systems (like, for instance,
the Aharonov--Bohm billiard~\cite{BR86} and the three-dimensional Anderson model at the
metal-insulator transition point~\cite{An58}) the spectrum statistics (in particular, the spacings
distribution) was found to be quite different from those of either chaotic or generic integrable
systems (see, e.g., Refs.~\cite{DJM95,GMW98}). In the late 90's, Bogomolny, Gerland and
Schmit~\cite{BGS99,BGS01} tried to account for this discrepancy by assuming that for the latter
systems the probability density~$p(\la_1,\dots,\la_N)$ of the eigenvalues~$\la_k$ (in a finite
range of the spectrum) is given by a nearest-neighbors version of the joint probability
distribution of the eigenvalues of the Gaussian $\be$-ensembles, namely\footnote{More precisely,
  the latter authors considered a periodic version of this density obtained by discarding the
  first factor, whose contribution is negligible in the limit $N\to\infty$, and adding an
  interaction term between the first and last particles.}
\[
   p(\la_1,\dots,\la_N)\propto\prod_{i=1}^N\e^{-\frac\be2\la_i^2}
  \cdot\prod_{i=1}^{N-1}|\la_i-\la_{i+1}|^\be\,.
\]
If we identify the eigenvalue~$\la_k$ with the coordinate~$x_k$ of a quantum particle, the above
distribution is the probability density of the ground state of the $N$-body Hamiltonian
\[
  H=-\sum_{i=1}^N\pdf{^2}{x_i^2}+\om\sum_{i=1}^Nx_i^2+\sum_{i=1}^{N-1}\frac{2\al(\al-1)}{(x_i-x_{i+1})^2}-\sum_{i=1}^{N-1}\frac{2\al^2}{(x_i-x_{i+1})(x_{i+1}-x_{i+2})},
\]
with~$2\al=2\om=\be$~\cite{JK99}. Note that, by contrast to the Calogero model, the latter
Hamiltonian features only nearest-neighbors (two-body) and next-to-nearest-neighbors (three--body)
interactions among the particles. Proceeding in a similar way with the joint probability density
of the eigenvalues of Dyson's circular ensembles one obtains a nearest-neighbors version of the
ground state of the Sutherland model, which is the ground state of a quantum many-body Hamiltonian
with trigonometric two- and three-body near-neighbors interactions. This connection between random
matrix theory and quantum many-body models with near-neighbors interactions of
Calogero--Sutherland type has in fact spurred the construction of further such models (including
particles with spin and interactions of arbitrary finite range) and the study of their properties
(see, e.g., Refs.~\cite{BK01,EFGR05b,EFGR07,TJK16,PBOC17}).

The purpose of this paper is to classify all quantum many-body models in one dimension with
nearest- and next-to-nearest-neighbors (translation invariant) interactions whose ground state
factorizes as in Eq.~\eqref{CSfactor}, but with the differences~$x_i-x_j$ replaced by the
nearest-neighbors differences $x_i-x_{i+1}$. As we have just remarked, these models include the
versions of the Calogero and Sutherland Hamiltonians with near-neighbors interactions introduced
in Ref.~\cite{JK99}. In other words, our goal is to perform the near-neighbors analog of the
well-known classification of CS-type models with a factorized ground state of the
form~\eqref{CSfactor}, started by Sutherland and Calogero and ultimately completed by Koprucki and
Wagner. We shall present the general solution of this classification problem, both for motion on
the real line and on a circle. By contrast with the corresponding problem featuring long-range
interactions, this general solution depends on an arbitrary function of one variable and (for
motion in the real line) an arbitrary positive parameter. Moreover, we shall show that the
three-body term appearing in all previously known examples is unavoidable. In other words, this
term must necessarily be present in any potential whose ground state is of the sought-for form. We
shall also see that the general solution contains an elliptic potential which yields in a suitable
limit the rational and trigonometric models introduced in Ref.~\cite{JK99}, as well as a new
hyperbolic model akin to the long-range one discussed by Forrester~\cite{Fo94jsp}.

We shall finish this Introduction with a brief outline of the paper's organization. In
Section~\ref{sec.main} we obtain a solution of the classification problem depending on an
arbitrary function, and prove that the three-body term that it contains cannot be expressed as an
external potential plus a two-body term. We show in Section~\ref{sec.uniq} that there is no other
solution, thus completing the proposed classification. Section~\ref{sec.GSC} is devoted to
verifying that the factorized eigenfunction associated with this solution is actually the
(square-integrable) ground state of the corresponding Hamiltonian, provided that the arbitrary
function which appears in the solution satisfies some natural physical requirements. In
Section~\ref{sec.examples} we discuss some particular models included in the general solution,
recovering the rational and trigonometric potentials of~Ref.~\cite{JK99} and introducing the new
hyperbolic and elliptic models mentioned above. The paper ends with a concluding section where we
summarize our results and indicate possible future developments.

\section{General solution}\label{sec.main}

We shall consider quantum many-body Hamiltonians of the form
\begin{equation}\label{Hdef}
  H=-\sum_{i=1}^N\pdf{^2}{x_i^2}+V(x_1,\dots,x_N)\,,%\equiv-\Delta+V(\bx)\,,
\end{equation}
where the potential
\begin{equation}\label{pot}
V(\bx)=\sum_i V_1(x_i)+\sum_i V_2(x_i-x_{i+1})+\sum_iV_3(x_i-x_{i+1},x_{i+1}-x_{i+2})
\end{equation}
features at most three-body near-neighbors translation invariant interactions. We shall assume
that the particles move either on a circle or on the real line. In the first case the coordinates
$x_i$ are typically angular variables, and the particles $1$ and $N$ are considered to be nearest
neighbors. In particular, in this case all sums and products will be assumed to run from $1$ to
$N$, with the identifications $x_{N+k}\equiv x_k$ for all $k\in\ZZ$. On the other hand, when the
particles move on the real line the coordinates $x_i$ are unbounded, and we shall not make the
latter identifications. Thus in this case all sums and products will be taken to run over the
largest meaningful range between $1$ and $N$. For instance, in this case
\[
  \sum_iV_1(x_i)\equiv\sum_{i=1}^NV_1(x_i)\,,\quad \sum_i
  V_2(x_i-x_{i+1})\equiv\sum_{i=1}^{N-1}V_2(x_i-x_{i+1})\,,
\]
and, in general,
\begin{equation}\label{sumcon}
  \sum_if(x_i,x_{i+1},\dots,x_{i+k})\equiv\sum_{i=1}^{N-k}f(x_i,x_{i+1},\dots,x_{i+k})\,.
\end{equation}
With this convention, we will be able to present the results for the cases of motion on
the circle or on the real line in a unified way.

Our goal is to classify the many-body potentials of the form~\eqref{pot}, both on a circle and on
the real line, for which~$H$ admits a ground state~$\psi(\bx)$ of the form
\begin{equation}\label{wave-func}
\psi(\bx)=\prod_i\rho(x_i)\cdot\prod_i\chi(x_i-x_{i+1})\,.
\end{equation}
Here $\rho$ and $\chi$ are two functions of one variable such that $\psi$ is square-integrable. In
other words (cf.~Eq.~\eqref{CSfactor}), the wave function~\eqref{wave-func} factorizes over the
(positive) \emph{simple} roots of the $A_{N-1}$ root system. We shall say that such a wave
function is \emph{Jastrow-like}, by analogy with the usual Jastrow-type form~\eqref{CSfactor}. We
will show that there is no solution featuring only two-body interactions, while the general
solution with three-body interactions depends on a constant and an arbitrary function of one
variable. The latter solution includes the well-known (rational and trigonometric) potentials of
Refs.~\cite{JK99,AJK01,EGKP05,EFGR05b,EFGR07} and their hyperbolic counterpart, as well as a new
elliptic potential which encompasses the previously known ones.

Imposing that~$\psi(\bx)$ in Eq.~\eqref{wave-func} be an eigenfunction of the
Hamiltonian~\eqref{Hdef} with energy~$E$ we readily obtain\goodbreak
\begin{multline}
  V(\bx)-E=\sum_i\big(\tau^2(x_i)+\tau'(x_i)\big)
                +2\sum_i\big(\vp^2(x_i-x_{i+1})+\vp'(x_i-x_{i+1})\big)\\
   +2\sum_i\vp(x_i-x_{i+1})\big(\tau(x_i)-\tau(x_{i+1})\big)
        -2\sum_i\vp(x_i-x_{i+1})\vp(x_{i+1}-x_{i+2}),
        \label{VmE}
      \end{multline}
where the one-variable functions $\tau$ and $\vp$ are by definition the logarithmic derivatives of
$\rho$ and $\chi$, i.e.,
\begin{equation}
\tau(x)\equiv\frac{\rho'(x)}{\rho(x)}\,,\qquad
\vp(x)\equiv\frac{\chi'(x)}{\chi(x)}\,.
\label{tausi}
\end{equation}
Here and in what follows we shall assume that $\vp'$ is not identically zero, since
otherwise~$\psi(\bx)$ is a product of one-particle states and there is no interaction between the
particles. We shall also assume, for simplicity's sake, that~$\tau$ and~$\vp$ are meromorphic
functions.

\begin{figure}[h]
  \centering
  \includegraphics[width=4cm]{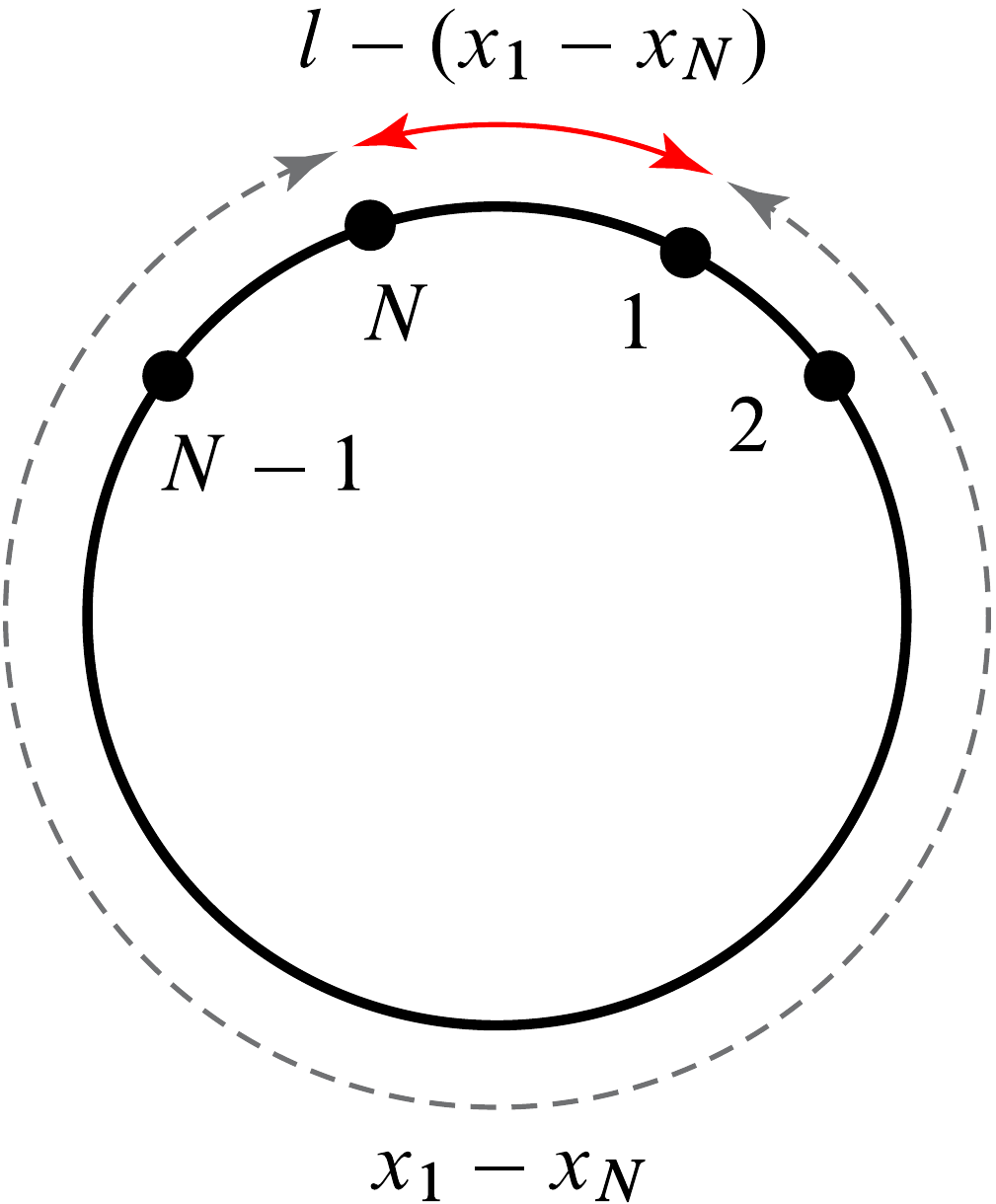}
  \caption{Arc distance~$l-(x_1-x_N)$ between the particles $1$ and~$N$ on a circle of
    circumference~$l$.}
  \label{fig.circle}
\end{figure}

It is readily apparent from Eq.~\eqref{VmE} that when~$\tau''\equiv0$ the potential~$V$ is already
of the sought for form~\eqref{pot}, both for the circle and the real line. Without loss of
generality (modulo a trivial overall translation of the coordinates), we can take
$\tau(x)=-\om x$. We thus obtain the following formulas for the potential $V$ and its Jastrow-like
eigenfunction~$\psi(\bx)$:
\begin{align}
   V(\bx)&=\om^2r^2+2\sum_i\big(\vp'(x_i-x_{i+1})+\vp^2(x_i-x_{i+1})\big)\notag\\
   &\hphantom{\om^2r^2}-2\om\sum_i(x_i-x_{i+1})\vp(x_i-x_{i+1})-2\sum_i\vp(x_i-x_{i+1})\vp(x_{i+1}-x_{i+2}),
        \label{Vfinal}\\
  \psi(\bx)&\propto\e^{-\om r^2/2}\prod_i\chi(x_i-x_{i+1})\,,\quad
       \chi(x)\equiv\exp\Big(\int^x\!\!\vp(s)\,\dd s\Big)\,,
              \label{psifinal}
\end{align}
with $r^2\equiv\sum_i x_i^2$ and energy $E=N\om$. Note that $\vp$ is an arbitrary function, except
for the requirement that the Jastrow-like eigenfunction~\eqref{psifinal} be square-integrable and
other restrictions that we shall discuss below. We emphasize that the previous result is valid
both for motion on the circle and on the line, with the convention for the range of the indices in
sums and products explained above. However, in the former case one should set $\om=0$, since the
external potential~$\om^2r^2$ is not periodic in the angular variables $x_i$. In addition, in both
cases there are several conditions that the function~$\vp$ should satisfy stemming from natural
physical requirements. In the first place, the two- and three-body potentials
\begin{equation}
  V_2(x)=2\big(\vp'(x)+\vp^2(x)-\om x\vp(x)\big)\,,\qquad V_3(x,y)=-2\vp(x)\vp(y)
  \label{V2V3gen}
\end{equation}
should be even functions of their arguments, i.e., $ V_2(-x)=V_2(x)$ and $V_3(-x,-y)=V_3(x,y)$\,.
This immediately implies that~$\vp(x)$ should be an odd function of~$x$. Furthermore, in the case
of motion in the circle the boundary terms in the second and third sums in Eq.~\eqref{pot} should
be consistent with the geometry of the system. For instance, the last term in the second sum,
given by
\[
  V_2(x_N-x_{N+1})\equiv V_2(x_N-x_1),
\]  
should be equal to~$V_2(l+x_N-x_1)$, since~$l-(x_1-x_N)$ is the (arc) distance between the
consecutive particles $N$ and~$1$ on a circle of circumference $l$ (cf.~Fig.~\ref{fig.circle}).
This implies that $V_2$ should be an~$l$-periodic function,
so that
\begin{equation}
  \label{V2per}
  V_2(x)=V_2(l+x)=V_2(l-x)\,,
\end{equation}
where the last equality is a consequence of the even character of~$V_2$. Similarly, from the last
two terms in the third sum of Eq.~\eqref{pot} we obtain the relations
\begin{align*}
  &V_3(x_{N-1}-x_N,x_N-x_{N+1})\equiv V_3(x_{N-1}-x_N,x_N-x_1)=V_3(x_{N-1}-x_N,l+x_N-x_1)\,,\\
  &V_3(x_N-x_{N+1},x_{N+1}-x_{N+2})\equiv V_3(x_N-x_1,x_1-x_2)=V_3(l+x_N-x_1,x_1-x_2),
\end{align*}
which lead to the periodicity conditions
\begin{equation}\label{V3per}
  V_3(x+l,y)=V_3(x,y+l)=V_3(x,y)\,.
\end{equation}
From Eq.~\eqref{V2V3gen} we easily see that conditions~\eqref{V2per}-\eqref{V3per} above are
equivalent to the relation
\begin{equation}\label{percond}
  \vp(x+l)=\vp(x)\,.
\end{equation}
In summary,~$\vp$ should be an odd function of its argument and, in the case of motion on a
circle, $l$-periodic. Thus in the latter case we have
\begin{equation}\label{pervp}
\vp(l-x)=\vp(-x)=-\vp(x)\,,
\end{equation}
so that~$\vp$ is odd about~$l/2$. Noting that\footnote{We cannot represent~$\chi$ by the following
  formula outside the open interval~$(0,l)$, since~$\vp$ is typically singular at~$x=0$, and hence
  at integer multiples of the period $l$. We shall assume in the following discussion that~$\vp$
  has no other singularities.}
\[
  \chi(x)=c\exp\bigg(\int_{l/2}^x\vp(t)\diff t\bigg),\qquad 0<x<l,
\]
(where $c$ is a constant) we deduce that~$\chi$ is symmetric about~$l/2$ in the interval~$(0,l)$,
i.e.,
\[
  \chi(x)=\chi(l-x)\,,\qquad 0<x<l\,.
\]
The latter formula and the periodicity conditions~\eqref{pervp} imply that
\[
  \chi(x)=\chi(l-x)=\chi(l+x)
\]
holds everywhere (except, at most, at integers multiples of~$l$). Thus in the case of motion on
the circle $\chi$ is an even, $l$-periodic function. In particular, it follows from
Eq.~\eqref{psifinal} that in this case the Jastrow-like wave function~$\psi$ is~$l$-periodic in
each variable.

To end this section, we shall next show that the solution~\eqref{Vfinal} does not include
potentials with purely two-body interactions. In other words, we must prove that the three-body
term in Eq.~\eqref{Vfinal} cannot be expressed as a sum of an external potential and a two-body
term, i.e., that the equation
\begin{equation}\label{twobodyeq}
  \sum_i\vp(x_i-x_{i+1})\vp(x_{i+1}-x_{i+2})=\sum_i\la(x_i)+\sum_iF(x_i-x_{i+1}),
\end{equation}
where $\la$, $F$ are functions of one variable, cannot be satisfied unless $\vp$ is constant. To
this end, consider first the case of motion on a circle, which is technically simpler due to the
symmetry under the cyclic group. In this case, using the elementary identities
\begin{align}\label{id1}
  &\sum_i\la(x_i)=\frac13\sum_i\big(\la(x_i)+\la(x_{i+1})+\la(x_{i+2})\big),\\
  &\sum_iF(x_i-x_{i+1})=\frac12\sum_i\big(F(x_i-x_{i+1})+F(x_{i+1}-x_{i+2})\big),
    \label{id2}
\end{align}
and calling $(x_i,x_{i+1},x_{i+2})\equiv(x,y,z)$ we deduce that Eq.~\eqref{twobodyeq} is
equivalent to the functional equation
\begin{equation}\label{func2bxyz}
  \frac13\big(\la(x)+\la(y)+\la(z)\big)+\frac12\big(F(x-y)+F(y-z)\big)=\vp(x-y)\vp(y-z).
\end{equation}
In terms of the independent variables~$x$, $u\equiv x-y$, $v\equiv y-z$, the latter equation can
be written as
\begin{equation}
  \frac13\big(\la(x)+\la(x-u)+\la(x-u-v)\big)+\frac12\big(F(u)+F(v)\big)=\vp(u)\vp(v)\,.
  \label{func2body}
\end{equation}
Differentiating with respect to~$x$ and setting $u=v=0$ we deduce that~$\la=\la_0$ is constant, so
that
\[
  \vp(u)\vp(v)=\frac12\big(F(u)+F(v)\big)+\la_0,
\]
and hence
\[
  \vp(u)^2=F(u)+\la_0\,.
\]
Substituting back in the previous equation we conclude that $\vp(u)-\vp(v)$ must be constant. This
implies that~$\vp$ itself is constant, which is excluded. In the case of motion on the line,
Eq.~\eqref{twobodyeq} still holds (with the convention~\eqref{sumcon} for the summation range),
but the identities~\eqref{id1}-\eqref{id2} should be replaced by
\begin{align}\label{id1l}
  &\sum_i\la(x_i)=\frac13\sum_i\big(\la(x_i)+\la(x_{i+1})+\la(x_{i+2})\big)\notag\\
  &\hphantom{\sum_i\la(x_i)=\frac13\sum_i\la(x_i)+}+\frac13\big(2\la(x_1)+\la(x_2)+\la(x_{N-1})+2\la(x_N)\big),\\
  &\sum_iF(x_i-x_{i+1})=\frac12\sum_i\big(F(x_i-x_{i+1})+F(x_{i+1}-x_{i+2})\big)\notag\\
  &\hphantom{\sum_i\la(x_i)=\frac13\sum_i\la(x_i)+}+\frac12\big(F(x_1-x_2)+F(x_{N-1}-x_N)\big).
    \label{id2l}
\end{align}
Consequently, in this case Eq.~\eqref{twobodyeq} is equivalent to Eq.~\eqref{func2bxyz} or
Eq.~\eqref{func2body}, together with the following two relations coming from the boundary terms in
Eqs.~\eqref{id1l}-\eqref{id2l}:
\[
  \frac13\big(2\la(x)+\la(y)\big)+\frac12 F(x-y)
  = -\frac13\big(\la(x)+2\la(y)\big)-\frac12 F(x-y)=c\,,
\]
where $c$ is a constant. Since we have just seen that Eq.~\eqref{func2body} cannot be satisfied
unless $\vp$ is a constant, we conclude that there are no potentials of the form~\eqref{Vfinal}
with only two-body interactions also in the case of motion on the line.

\section{Uniqueness}\label{sec.uniq}

We shall show in this section that~\eqref{Vfinal} is the most general potential of the
form~\eqref{pot} admitting a Jastrow-like eigenfunction~\eqref{wave-func}. Together with the
result at the end of the previous section, this implies that no potential of the form~\eqref{pot}
with only two-body interactions admits a Jastrow-like eigenfunction~\eqref{wave-func}. For
clarity's sake, we shall deal separately with the case of motion on a circle and on the real line.

\subsection{Motion on a circle}
To begin with, note that in this case we can express
the sum of a two- and a three-body term as a pure three-body term, namely
\begin{equation}\label{F23}
  \sum_iF_2(x_i-x_{i+1})+\sum_iF_3(x_i-x_{i+1},x_{i+1}-x_{i+2})=\sum_iF(x_i-x_{i+1},x_{i+1}-x_{i+2})\,,
\end{equation}
with $F(x,y)=F_3(x,y)+(F_2(x)+F_2(y))/2$. Hence the RHS of Eq.~\eqref{VmE} will be of the
form~\eqref{pot} provided that there exist a function $\la$ of one variable and a function $F$ of
two variables such that
\begin{equation}\label{laF}
 \sum_i\la(x_i)+\sum_i F(x_i-x_{i+1},x_{i+1}-x_{i+2})=2\sum_i\vp(x_i-x_{i+1})\big(\tau(x_i)-\tau(x_{i+1})\big).
\end{equation}
Taking into account Eq.~\eqref{id1} and the analogous identity
\begin{multline*}
 2\sum_i\vp(x_i-x_{i+1})\big(\tau(x_i)-\tau(x_{i+1})\big)\\
 =\sum_i\vp(x_i-x_{i+1})\big(\tau(x_i)-\tau(x_{i+1})\big)
+\sum_i\vp(x_{i+1}-x_{i+2})\big(\tau(x_{i+1})-\tau(x_{i+2})\big),
\end{multline*}
and calling again $(x_i,x_{i+1},x_{i+2})\equiv(x,y,z)$, we arrive at the functional equation
\begin{multline}
   \frac13\big(\la(x)+\la(y)+\la(z)\big)
  +F(x-y,y-z)\\
  =\vp(x-y)\big(\tau(x)-\tau(y)\big)+\vp(y-z)\big(\tau(y)-\tau(z)\big).
                                         \label{3eqxyz}
\end{multline}
Equivalently, setting $u\equiv x-y$ and $v\equiv y-z$ we can rewrite the latter equation as
\begin{equation}
  \label{L3eqxuv}
  L(x,u,v)=F(u,v)\,,
\end{equation}
where the function $L(x,u,v)$ is defined as
\begin{multline}\label{Leq}
 L(x,u,v)\equiv\vp(u)\big(\tau(x)-\tau(x-u)\big)
+\vp(v)\big(\tau(x-u)-\tau(x-u-v)\big)\\
-\frac13\big(\la(x)+\la(x-u)+\la(x-u-v)\big).
\end{multline}
In particular, from the last two equations it easily follows that~$\la$ and~$F$ are meromorphic
functions of their arguments.

We shall now show that Eq.~\eqref{L3eqxuv} implies that $\tau''\equiv0$, which, as explained in
the previous section, yields the potential~\eqref{Vfinal}. The key idea in our proof is to note
that by Eq.~\eqref{L3eqxuv} the partial derivative of~$L(x,u,v)$ with respect to~$x$ must vanish
identically, i.e.,
\begin{multline}
  \frac13\big(\la'(x)+\la'(x-u)+\la'(x-u-v)\big)\\
  =\vp(u)\big(\tau'(x)-\tau'(x-u)\big)+\vp(v)\big(\tau'(x-u)-\tau'(x-u-v)\big).
  \label{deq}
\end{multline}
Letting $v\to-u$ in the latter equation and taking into account the odd character of~$\vp$
we obtain
\[
2\la'(x)+\la'(x-u)=6\,\vp(u)\big(\tau'(x)-\tau'(x-u)\big).
\]
If $\tau''(x)\not\equiv0$ we can solve for~$\vp(u)$ in the latter equation, with the result
\begin{equation}
  \label{dequ}
  6\,\vp(u)=\frac{2\la'(x)+\la'(x-u)}{\tau'(x)-\tau'(x-u)}\,.
\end{equation}
Expanding the RHS of this equality in a Laurent series around~$u=0$ we readily obtain
\[
  6\,\vp(u)=\frac{3\la'(x)}{\tau''(x)u}+\frac{3\la'(x)\tau'''(x)}{2\tau''(x)^2}
  -\frac{\la''(x)}{\tau''(x)}+\Or(u)\,.
\]
The coefficient of~$1/u$ in the latter equation must be a constant~$6\al$, while that of~$u^0$
must vanish on account of the odd character of~$\vp$. We thus deduce that
\[
  \la'(x)=2\al\tau''(x)\,,\qquad \al\tau'''(x)=0\,.
\]
From the latter equations it also follows that~$\al\ne0$, since otherwise~$\vp$ would vanish
identically on account of~Eq.~\eqref{dequ}. Hence $\tau'''=0$, so that can write
\[
  \tau(x)=\tau_0+\tau_1x+\tau_2x^2\,,\qquad \la(x)=\la_0+4\al\tau_2x\,,
\]
with $\la_0$, $\tau_i$ constant and~$\tau_2\ne0$. Substituting into Eq.~\eqref{dequ} we easily
obtain
\[
  \vp(u)=\frac\al u\,.
\]
However, this solution is not acceptable, since it does not satisfy the periodicity
condition~\eqref{percond} that should hold in this case. We thus conclude that in the case of
motion on the circle there is no solution of the problem posed with $\tau''\not\equiv0$, as
claimed.

\subsection{Motion on the real line}

We shall next discuss the case of motion on the real line, in which the coordinates $x_i$ are
unbounded. To begin with, in this case the identity~\eqref{F23} should be replaced by
\begin{multline}
   \sum_iF_2(x_i-x_{i+1})+\sum_iF_3(x_i-x_{i+1},x_{i+1}-x_{i+2})\\
  =\sum_iF(x_i-x_{i+1},x_{i+1}-x_{i+2})+\frac12\big(F_2(x_1-x_2)+F_2(x_{N-1}-x_N)\big),
\end{multline}
where as before $F(x,y)=F_3(x,y)+(F_2(x)+F_2(y))/2$, and we are using the
convention~\eqref{sumcon} on the range of summation indices. Consequently, Eq.~\eqref{laF} now
reads
\begin{multline}
   \sum_i\la(x_i)+\sum_i F(x_i-x_{i+1},x_{i+1}-x_{i+2})+
  G(x_1-x_2)+G(x_{N-1}-x_N)\\
  =2\sum_i\vp(x_i-x_{i+1})\big(\tau(x_i)-\tau(x_{i+1})\big),
  \label{laFG}
\end{multline}
where~$\la$, $G$ are functions of one variable and $F$ is a function of two variables. Using
Eq.~\eqref{id1l} and the
identity
\begin{align*}
  2\sum_i&\vp(x_i-x_{i+1})\big(\tau(x_i)-\tau(x_{i+1})\big)\\&=
  \vp(x_1-x_2)\big(\tau(x_1)-\tau(x_{2})\big)+\vp(x_{N-1}-x_N)\big(\tau(x_{N-1})-\tau(x_N)\big)\\
         &\hphantom{={}}+\sum_i\Big[\vp(x_i-x_{i+1})\big(\tau(x_i)-\tau(x_{i+1})\big)
           +\vp(x_{i+1}-x_{i+2})\big(\tau(x_{i+1})-\tau(x_{i+2})\big)\Big]
\end{align*}
in Eq.~\eqref{laFG} we readily obtain Eq.~\eqref{3eqxyz}, or
equivalently~\eqref{L3eqxuv}-\eqref{Leq}, plus the additional constraints
\begin{align}
  \big(\tau(x)-\tau(y)\big)\vp(x-y)&=\frac13\big(2\la(x)+\la(y)\big)+G(x-y)+c\notag\\
                                   &=\frac13\big(\la(x)+2\la(y)\big)+G(x-y)-c\,,
  \label{constraints}
\end{align}
where $c$ is a constant. From the latter equations we immediately deduce that~$\la=\la_0$ must be
a constant. This in turn implies that~$\tau(x)-\tau(y)$ is a function of $x-y$, so that $\tau$
must be linear in~$x$. Thus also in this case there is no solution with $\tau''\not\equiv0$, as
claimed.

\section{Ground state conditions}\label{sec.GSC}
We shall show in this section that the Jastrow-like eigenfunction~\eqref{psifinal} is actually the
ground state of the potential~\eqref{Vfinal}, provided that the function $\vp$ satisfy very
general assumptions that we shall now discuss.

We shall start our discussion with the case of motion on the line. First of all, it is natural on
physical grounds to require that~$\vp$ be analytic everywhere except at the origin, so that the
only singularities of the potential~\eqref{Vfinal} are located on the hyperplanes~$x_i=x_{i+1}$.
For simplicity, we shall further assume that~$\vp$ has a simple pole at the origin, i.e.,
\[
\vp(x)=\frac\al x+\vp_0(x)\,,
\]
with~$\al\ne0$ and $\vp_0$ analytic on the real line. The wave function~\eqref{psifinal} can thus
be written as
\begin{equation}\label{psiform}
\psi(\bx)\propto\e^{-\om r^2/2}\prod_i|x_i-x_{i+1}|^\al\cdot\prod_i\e^{\Phi(x_i-x_{i+1})}\,,
\end{equation}
with~$\Phi(x)=\int^x\vp_0(s)\diff s$ analytic on the real line. Note that we must have~$\al>1/2$, to
ensure that the expected value of the kinetic energy of the eigenstate~$\psi$ be finite. This
automatically guarantees the square integrability of~$\psi$ near the singular hyperplanes
$x_i=x_{i+1}$. Moreover, since the potential~\eqref{Vfinal} diverges near these hyperplanes as
$\al(\al-1)(x_i-x_{i+1})^{-2}$, if~$\al\ne1$ the particles cannot overtake each
other~\cite{Ca71,An-AJP76}. Thus we can fix the ordering of the particles as, e.g.,
$x_1>\cdots>x_N$, which amounts to taking the configuration space of the system as the open set
\[
  A= \{\bx\in\RR^N\mid x_1>\cdots> x_N\}.
\]
It is then clear that~$\psi$ does not vanish on~$A$ by Eq.~\eqref{psiform}. Hence to show
that~$\psi$ is indeed the ground state of $H$ it suffices to verify that it is square-integrable
at infinity. To this end, we need only impose that\footnote{If $\om=0$, the
  potential~\eqref{Vfinal} is translation-invariant, so that the total momentum is conserved. In
  this case we can separate the center of mass motion and regard the differences~$x_i-x_{i+1}$,
  $1\le i\le N-1$, as independent variables. Consequently, the eigenfunction~\eqref{wave-func}
  with~$\om=0$ will be square-integrable at infinity provided that $\Phi(x)\le -c_0\log|x|+c_1$
  with $c_0>1$.} $\om>0$ and~$\Phi(x)\le cx^2$ with $c<\om/8$. Indeed, if this is the case we have
\begin{equation}\label{B}
  \e^{-\om
    r^2/2}\prod_i\e^{\Phi(x_i-x_{i+1})}\le\exp\Big(-\tfrac12\,\om r^2
  +c\sum_i(x_i-x_{i+1})^2\Big)\equiv\e^{-\tfrac12\sum_{i,j}b_{ij}x_ix_j}\,,
\end{equation}
where~$B\equiv (b_{ij})_{1\le i,j\le N}$ is the circulant matrix with first row
$(\om-4c,2c,\dots,2c)$ (the dots standing for zeros)~\cite{Gr06}. Since the eigenvalues of~$B$,
given by\footnote{For~$N=2$, the eigenvalues of~$B$ are $\om-2c$ and~$\om-6c$.}
\[
  \la_j=\om-8c\sin^2(j\pi/N)\,,\qquad j=0,\dots,N-1\,,
\]
are all positive on account of the condition~$\om>8c$, the associated quadratic form is positive
definite. By the inequality~\eqref{B}, this implies that the eigenfunction~\eqref{psiform} is
square-integrable at infinity.

In the case of motion on a circle, we shall again require that~$\vp$ have a simple pole at the
origin with residue $\al\ne0$. By the periodicity condition~\eqref{percond}, $\vp$ must have
simple poles with residue~$\al$ at integer multiples of the circle's circumference~$l$, so that
\[
  \vp(x)=\al\bigg(\frac1x+\frac1{x-l}\bigg)+\vp_0(x),
\]
with $\vp_0$ analytic on the interval~$[0,l]$. The Jastrow-like eigenfunction~\eqref{wave-func}
(with $\om=0$) is then given by
\begin{equation}
  \label{psicirc}
  \psi(\bx)\propto\prod_i\big|(x_i-x_{i+1})(l-x_i+x_{i+1})\big|^\al\cdot\prod_i\e^{\Phi(x_i-x_{i+1})},
\end{equation}
where~$\Phi(x)=\int^x\vp_0(s)\diff s$ is analytic on the interval~$[0,l]$. As before, the square
integrability of~the eigenfunction~$\psi$ at~$x_i-x_{i+1}=kl$ (with~$k\in\ZZ$) and the finiteness
of the average kinetic energy require that $\al>1/2$. Furthermore, the potential~\eqref{Vfinal}
diverges near the singular hyperplanes~$x_i-x_{i+1}=kl$ as~$\al(\al-1)(x_i-x_{i+1}-kl)^{-2}$.
Hence if $\al\ne1$ the particles cannot overtake each other, and the system's configuration space
can thus be taken as the open set
\begin{equation}\label{A}
A=\{\bx\in\RR^N\mid x_1>\cdots>x_N>x_1-l\}\,.
\end{equation}
The Jastrow-like eigenfunction~$\psi$ in Eq.~\eqref{psicirc} is square-integrable on $A$, since
the potential is translation-invariant and we can therefore regard the differences $x_i-x_{i+1}$,
which range over the bounded interval~$(0,l)$, as independent variables after separating the
center of mass motion. Moreover, $\psi$ does not vanish on the configuration space~\eqref{A}, and
is thus again the system's ground state.

\section{Examples}\label{sec.examples}

As we have seen in the previous sections, the most general potential of the form~\eqref{pot}
admitting a Jastrow-like eigenfunction~\eqref{psifinal} depends on an essentially arbitrary
function~$\vp$ of one variable and, in the case of motion on the line, an additional
constant~$\om$. In particular, choosing $\vp$ appropriately one should be able to recover all the
potentials of the form~\eqref{pot} previously proposed in the literature, as well as several
interesting generalizations thereof. Thus, if $\vp(x)=\al/x$, from Eqs.~\eqref{Vfinal}
and~\eqref{psifinal} we obtain the rational potential introduced in Ref.~\cite{JK99}, namely
\begin{align*}
  V(\bx)&=\om^2r^2+\sum_i\frac{2\al(\al-1)}{(x_i-x_{i+1})^2}-\sum_i\frac{2\al^2}{(x_i-x_{i+1})(x_{i+1}-x_{i+2})},\\%\label{Vrat}
  \psi(\bx)&\propto \e^{-\om r^2/2}\prod_i|x_i-x_{i+1}|^\al\,,
             \qquad E=N\om+2(N-1)\al\om\,.%\label{psirat}
\end{align*}
Although the previous formulas for~$V$ and~$\psi$ are \emph{formally} valid both for the circle
and the real line, as discussed in Section~\ref{sec.main} the above potential has a natural
physical interpretation only in the latter case\footnote{Note, however, that the variant of the
  latter potential with cyclic symmetry and its spin version have been used in Ref.~\cite{EFGR08}
  to construct an analogue of the Polychronakos--Frahm spin chain~\cite{Po93,Fr93} with
  nearest-neighbors interactions, whose first few eigenvalues can be computed in closed form.}.

Similarly, the choice $\vp(x)=(\pi\al/l)\cot(\pi x/l)$ with $\om=0$ leads to the trigonometric
potential~\cite{JK99}
\begin{align*}
   V(\bx)&=\big(\tfrac\pi l\big)^2\sum_i\frac{2\al(\al-1)}{\sin^2\bigl(\tfrac \pi l(x_i-x_{i+1})\bigr)}-2\big(\tfrac{\pi\al}l\big)^2
              \sum_i\cot\bigl(\tfrac \pi l(x_i-x_{i+1})\bigr)\cot\bigl(\tfrac \pi l(x_{i+1}-x_{i+2})\bigr),\\%\label{Vtrig}
  \psi(\bx)&\propto\prod_i\big|\sin\bigl(\tfrac \pi l(x_i-x_{i+1})\bigr)\big|^\al\,,\qquad
                E=2N\Big(\frac{\pi\al}l\Big)^{\!2}\,.%
  % \label{psitrig}
\end{align*}
The natural interpretation of this model is on a circle of radius $l/(2\pi)$.

A hyperbolic version of the previous potential is easily obtained by taking $l=\iu\pi/\be$ (with
$\be>0$) in the previous formula for~$\vp(x)$. We thus obtain $\vp(x)=\al\be\coth(\be x)$ and
\begin{align}
   V(\bx)&=\om^2r^2-2\al\be\om\sum_i(x_i-x_{i+1})\coth\bigl(\be(x_i-x_{i+1})\bigr)
              +\sum_i\frac{2\al(\al-1)\be^2}{\sinh^2\bigl(\be(x_i-x_{i+1})\bigr)}\notag\\
   &\hphantom{\om^2r^2-{}}-2\al^2\be^2\sum_i\coth\bigl(\be(x_i-x_{i+1})\bigr)\coth\bigl(\be(x_{i+1}-x_{i+2})\bigr),\label{Vhyp}\\
   \psi(\bx)&\propto\e^{-\om r^2/2}\prod_i\big|\sinh\bigl(\be(x_i-x_{i+1})\bigr)\big|^\al\,,\qquad
              E=N\om-2(N-1)\al^2\be^2\,.
              \label{psihyp}
\end{align}
Note that we have taken $\om>0$, since the latter potential has a natural physical interpretation
only on the line, and the term $\e^{-\om r^2/2}$ in the expression for $\psi$ is therefore needed
to guarantee its square integrability. The hyperbolic model~\eqref{Vhyp} can be regarded as the
near-neighbors analogue of the long-range model of CS type introduced by Forrester~\cite{Fo94jsp}.
The ground state of the latter model, which is similar to~\eqref{psihyp} but is factorized over
the whole~$A_{N-1}$ root system, was shown by Forrester to describe a Wigner solid in the
thermodynamic limit.

\begin{figure}[h]
  %\centering
  \includegraphics[height=.34\textwidth]{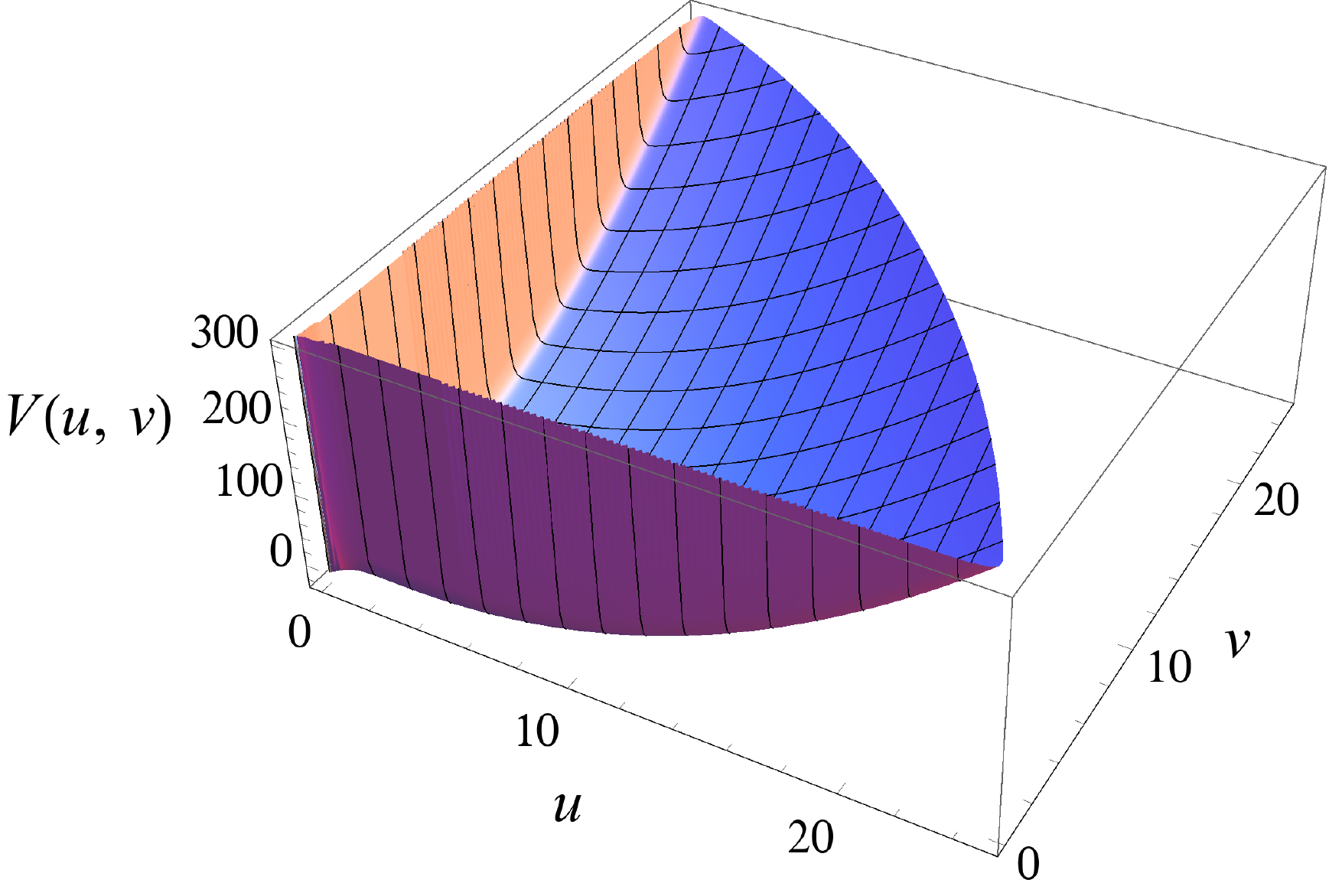}\hfill
  \includegraphics[height=.34\textwidth]{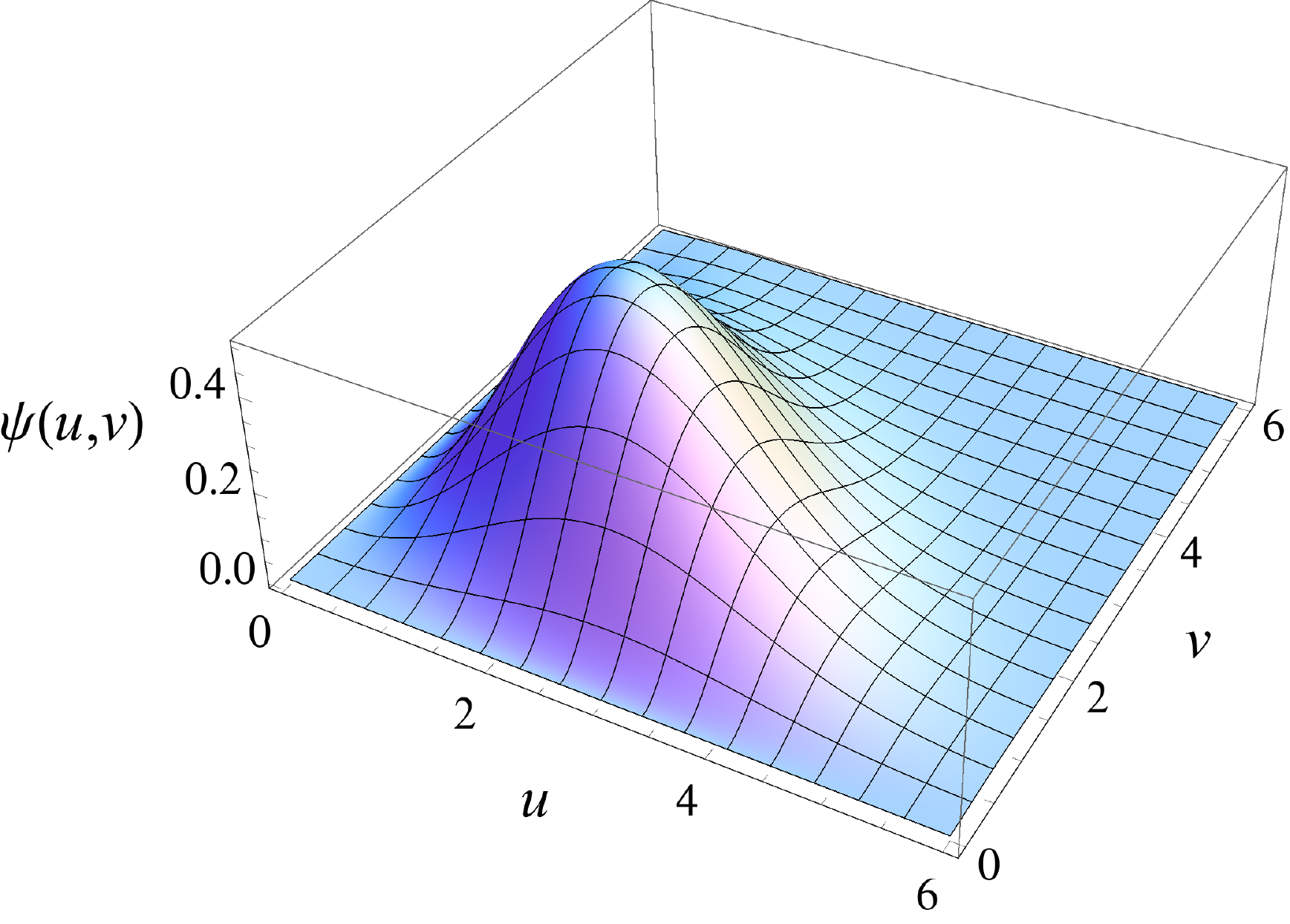}
  \caption{Hyperbolic potential~\eqref{Vhyp} (left) and its normalized Jastrow-like
    eigenfunction~\eqref{psihyp} (right) for $N=3$, $\al=2$, $\om=\be=1$ in the center of mass
    frame $x_1+x_2+x_3=0$ as a function of the relative coordinates $u\equiv x_1-x_2$,
    $v\equiv x_2-x_3$.}
  \label{fig.V3psihyp}
\end{figure}

\begin{figure}[h]
  %\centering
  \includegraphics[height=.34\textwidth]{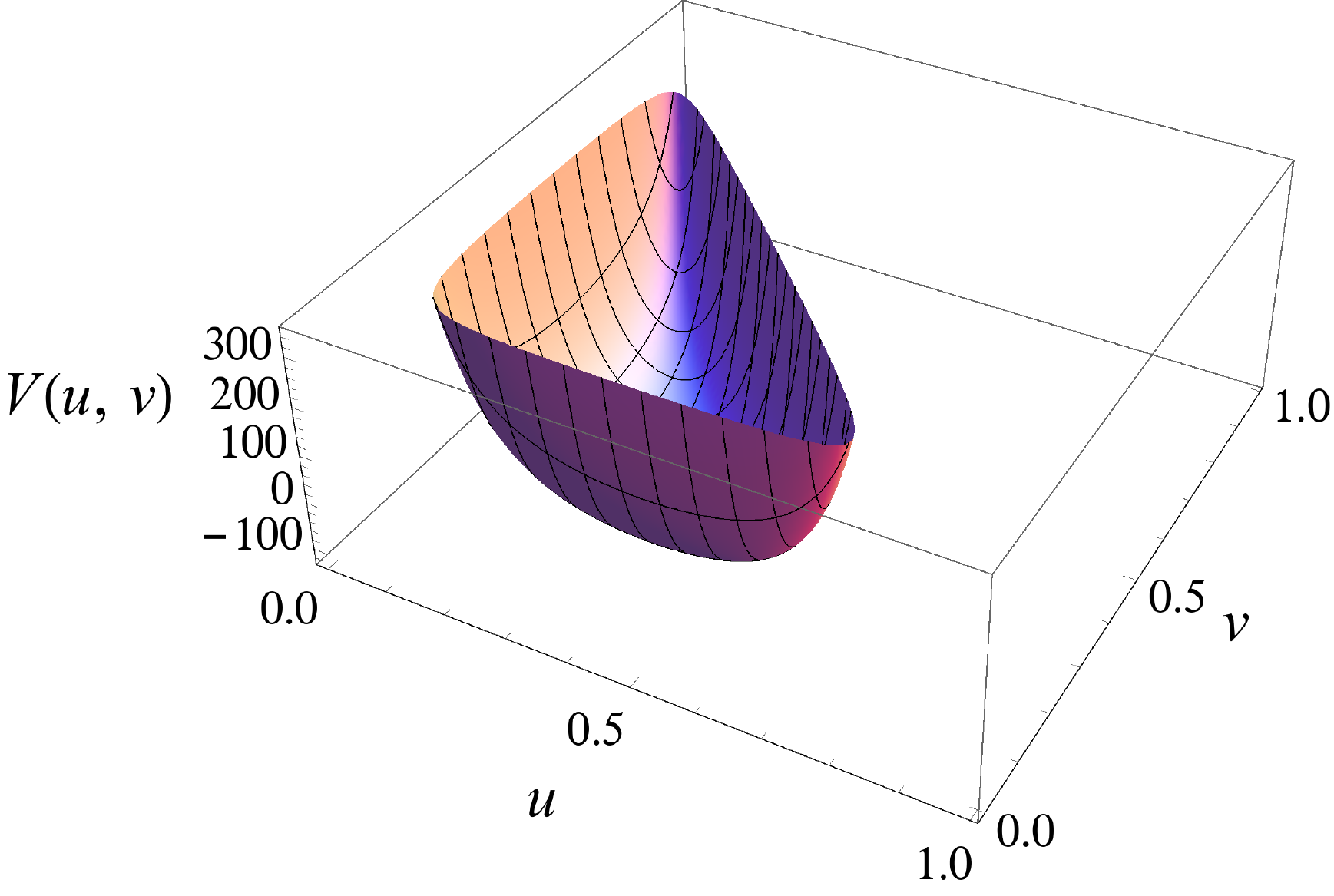}\hfill
  \includegraphics[height=.34\textwidth]{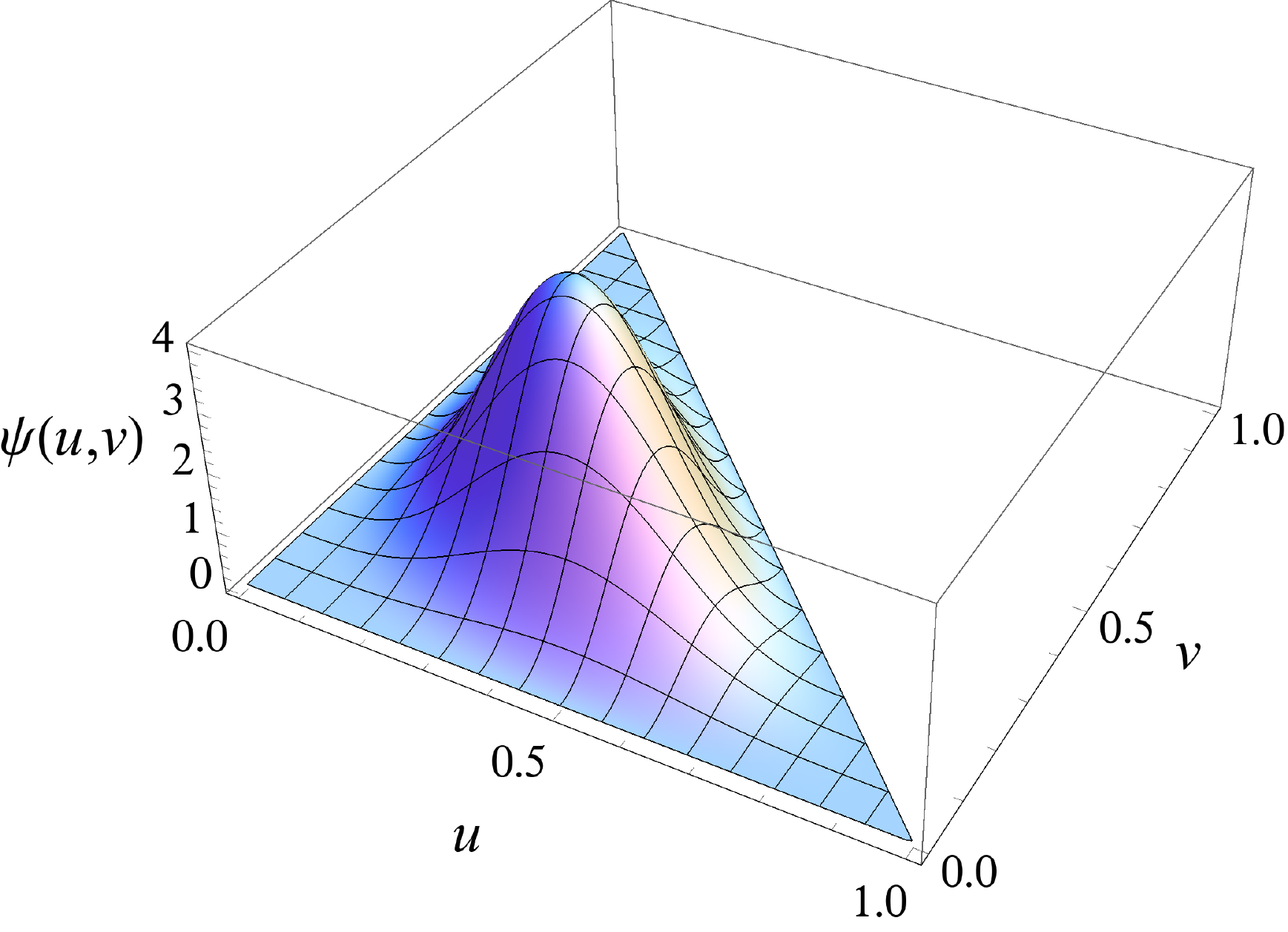}
  \caption{Elliptic potential~\eqref{Vell} (left) and its normalized Jastrow-like
    eigenfunction~\eqref{psiell} (right) for $N=3$, $\al=2$, $l=1$ and~$\Im \om_3=1/2$ as a
    function of the relative coordinates $u\equiv x_1-x_2$, $v\equiv x_2-x_3$.}
  \label{fig.V3psi}
\end{figure}

\begin{figure}[h]
  \includegraphics[width=.49\textwidth]{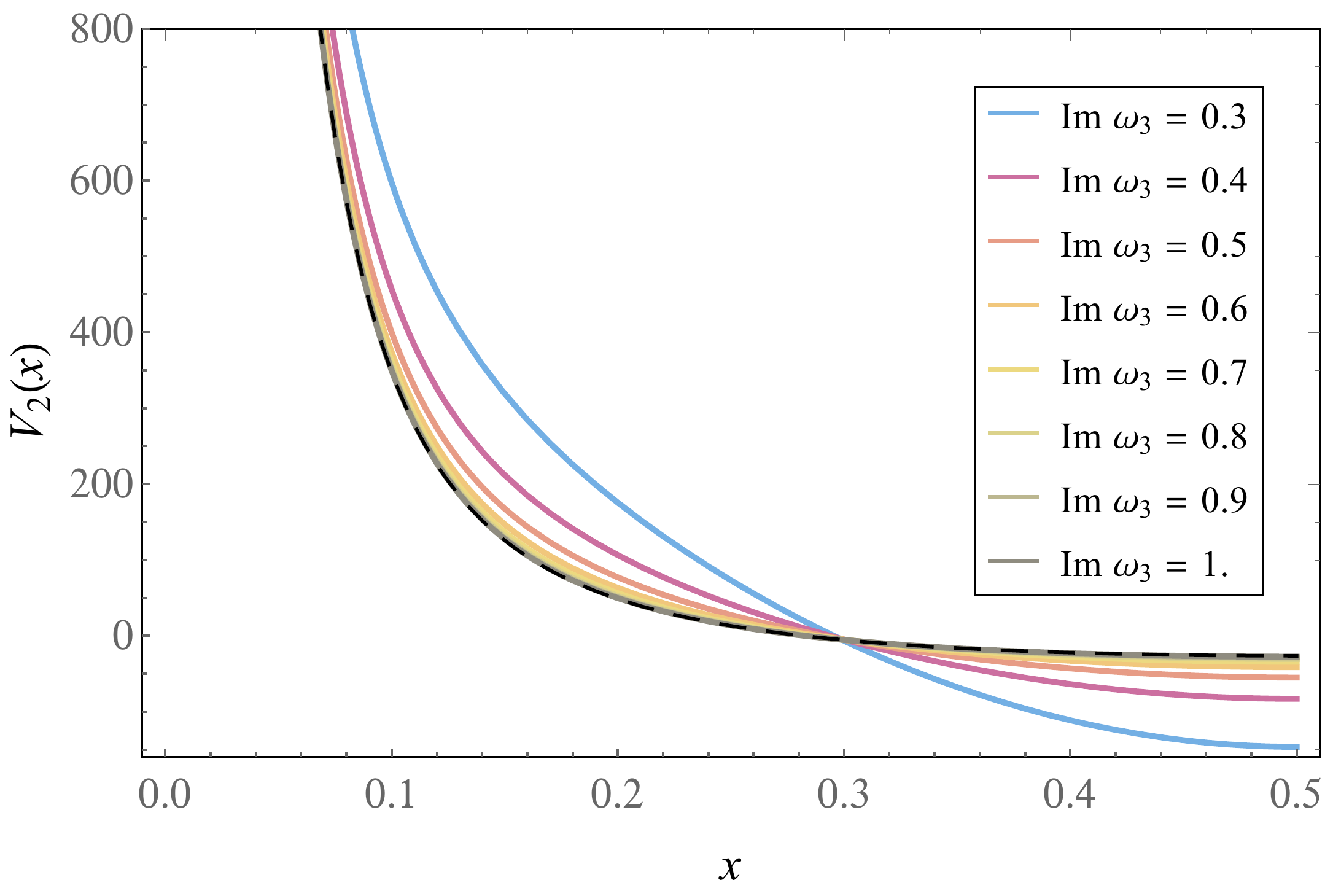}\hfill
  \includegraphics[width=.49\textwidth]{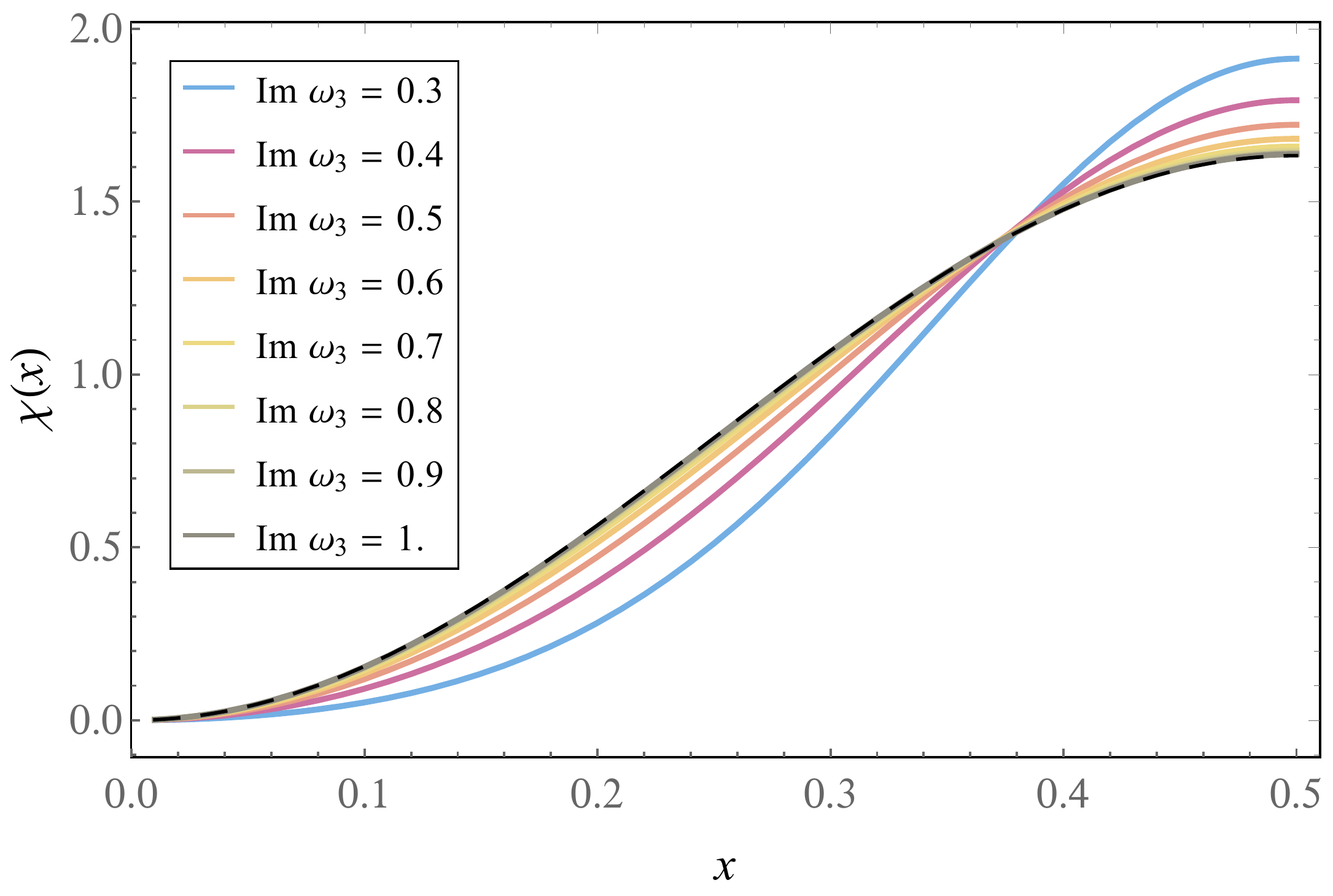}\hfill
  \caption{Left: elliptic two-body potential~\eqref{V2trig} with $l=1$, $\al=2$ and several values
    of $\Im\om_3$, compared to its limiting trigonometric potential~$4\sin^{-2}(\pi x)-20\pi^2/3$
    (dashed black line). Right: analogous plot for the corresponding functions~$\chi(x)$
    determining the Jastrow-like eigenfunction~\eqref{psifinal} (with the normalization
    $\int_0^1|\chi(x)|^2\diff x=1$).}
  \label{fig.V2chi}
\end{figure}

In the classification of long-range interaction potentials with two-body interactions and
Jastrow-type ground state performed in Refs.~\cite{Ca75,IM84,KW00}, the rational, trigonometric
and hyperbolic solutions are obtained precisely from the three choices of the function~$\vp$ used
in the previous examples. In fact, in this case there is an additional solution given by
\begin{equation}\label{ellgen}
  \vp(x)=\al\ze(x)+\ga x\,,
\end{equation}
where~$\ga\in\RR$ and~$\ze(x)\equiv\ze(x;g_2,g_3)$ is the Weierstrass
zeta function with invariants $g_2$ and $g_3$~\cite{WW27}, which yields the above three choices
of~$\vp$ as particular cases on account of the identities
\[
  \left\{
  \begin{aligned}
  &\ze(x;\tfrac43\be^4,\tfrac8{27}\be^6)-\frac{\be^2x}3=\be\cot(\be x)\,,\\
  &\ze(x;0,0)=\frac1x\,,\\
  &\ze(x;\tfrac43\be^4,-\tfrac8{27}\be^6)+\frac{\be^2x}3=\be\coth(\be x)\,.
\end{aligned}
\right.
\]
It is therefore natural to consider the potential~\eqref{Vfinal} generated by the function~$\vp$
in Eq.~\eqref{ellgen}. We shall assume that the invariants $g_{2,3}$ are real and satisfy the
condition~$g_2^3> 27g_3^2$, so that $\ze$ is real for real values of its argument and the
corresponding Weierstrass function~$\wp(z)\equiv-\ze'(z)$ has a real fundamental period~$l<\infty$
and a purely imaginary one~$\om_3$ (with~$\Im\om_3>0$). Since the $\ze$ function has simple poles
at integer multiples of these periods, the corresponding potential~\eqref{Vfinal} is naturally
defined on a circle of circumference~$l$. As explained in Section~\ref{sec.main}, this requires
that~$\om=0$ and that $\vp$ be an $l$-periodic function. In view of the identity
\[
  \ze(z+l)=\ze(z)+2\eta_1,
\]
where $\eta_1\equiv\ze(l/2)$, the latter condition will be satisfied if and only
if~$\ga=-2\al\eta_1/l$. We are thus led to consider the choice
\begin{equation}\label{vpell}
  \vp(x)=\al\bigg(\ze(x)-\frac{2\eta_1}{l}x\bigg)\,,
\end{equation}
whose associated potential is  given by
\begin{multline}
  V(\bx)=-2\al\sum_i\wp(x_i-x_{i+1})
  +2\sum_i\vp(x_i-x_{i+1})^2\\
  -2\sum_i\vp(x_i-x_{i+1})\vp(x_{i+1}-x_{i+2})
  \label{Vell}
\end{multline}
with~$\al>1/2$. The corresponding Jastrow-like eigenfunction and energy read
\begin{equation}
  \label{psiell}
  \psi(\bx)\propto\exp\Big(-\frac{\al\eta_1}{l}\sum_i(x_i-x_{i+1})^2\Big)
\prod_i|\si(x_i-x_{i+1})|^\al\,,\qquad E=\frac{4N\eta_1\al}{l},
\end{equation}
where the Weierstrass~$\si$ function is defined by $\si'/\si=\ze$ and~$\lim_{z\to0}\si(z)/z=1$
(see Fig.~\ref{fig.V3psi} for a plot of the potential~\eqref{Vell} and its Jastrow-like
eigenfunction~\eqref{psiell} for $N=3$ particles when~$\al=2$, $l=1$ and~$\Im\om_3=1/2$). Recall
that~$\si$ is entire and odd, and it vanishes only at the periods of~$\wp$, so that in
particular~$\si(kl)=0$ for all~$k\in\ZZ$. Hence~$\psi$ does not vanish on the configuration
space~\eqref{A}, and is therefore the system's ground state. We also know from the general
discussion of Section~\ref{sec.main} (and is also obvious from the $l$-periodicity of~$\wp$
and~$\vp$) that the two-body potential
\begin{equation}\label{V2trig}
  V_2(x)=2\big(-\al\wp(x)+\vp(x)^2\big)
\end{equation}
is~$l$-periodic and symmetric about~$l/2$
(cf.~Eq.~\eqref{V2per}), and that the Jastrow-like eigenfunction~\eqref{psiell} is also
$l$-periodic in each of its variables. The latter fact can also be checked directly with the help
of the identity
\[
  \si(z+l)=-\e^{2\eta_1(z+\frac l2)}\si(z)\,.
\]
In Fig.~\ref{fig.V2chi} we present a plot of $V_2$ and~$\chi$ for $\al=2$, $l=1$ and several
values of~$\Im\om_3$ in the half-period~$0<x<1/2$.

The potential~\eqref{Vell} depends on three real parameters, namely~$\al>1/2$, $\Im\om_3>0$,
and~$l>0$. Note, however, that from the well known identities
\[
  \wp(\mu x|\mu l/2,\mu\om_3)=\frac1{\mu^2}\wp(x)\,,\quad
  \ze(\mu x|\mu l/2,\mu\om_3)=\frac1{\mu}\ze(x)
\]
(where~$f(z|\mu l/2,\mu\om_3)$ denotes the corresponding Weierstrass function~$f$ with
periods~$\mu l$ and~$2\mu\om_3$) it easily follows that either~$\Im\om_3$ or $l$ can be rescaled to
(say)~$1$ by an appropriate overall dilation of the coordinates. Note also that when
$\Im\om_3\to\infty$ we have
\[
  \wp(x)\to\frac{\pi^2}{l^2}\bigg(\sin^{-2}\bigl(\tfrac{\pi x}l\bigr)-\frac13\bigg)\,,\quad
  \ze(x)\to\frac{\pi^2x}{3l^2}+\frac{\pi}l\cot\bigl(\tfrac{\pi x}l\bigr)\,,\quad
  \eta_1\to\frac{\pi^2}{6l}
\]
(see, e.g., Refs.~\cite{OLBC10,FG14JSTAT}), and consequently
\[
  \vp(x)\to\frac{\al\pi}l\cot\bigl(\tfrac{\pi x}l\bigr)\,,\qquad
  V_2(x)\to\frac{2\al(\al-1)}{\sin^2\bigl(\tfrac{\pi
      x}l\bigr)}-2\left(\frac{\pi}l\right)^2\!\al\bigg(\al-\frac13\bigg)
\]
(cf.~Fig.~\ref{fig.V2chi}). From these equations it readily follows that as~$\Im\om_3\to\infty$
the potential~\eqref{Vell} tends to
\[
  V_{\text{trig}}(\bx)-2\left(\frac{\pi}l\right)^2\!N\al\bigg(\al-\frac13\bigg),
\]
where~$V_{\text{trig}}(\bx)$ is the trigonometric potential in Ref.~\cite{JK99} discussed above.

\section{Summary and outlook}

In this paper we completely solve the problem of classifying all one-dimensional quantum
Hamiltonians with nearest- and next-to-nearest-neighbors (translation invariant) interactions
admitting a Jastrow-like ground state, both for motion on the real line and on a circle. This is
the simplest near-neighbors analogue of the well-known problem for Calogero--Sutherland models
with long-range interactions proposed shortly after their introduction and completely solved in
Ref.~\cite{KW00}. Our solution differs in two fundamental ways with its long-range counterpart. In
the first place, we show that the potential must necessarily contain a three-body interaction
term, which by construction is absent in the long-range solution. Secondly, the near-neighbors
solution depends on an essentially arbitrary function of one variable (and, for motion on the
line, on an additional positive parameter). The general solution contains a potential featuring
elliptic interactions, which yields the (rational and trigonometric) particular solutions
considered so far~\cite{JK99} as limiting cases.

Our results suggest several lines of work for further research. To begin with, it would certainly
be of interest to study in detail the potentials contained in the general solution, and in
particular determine whether one can exactly compute other eigenfunctions besides the ground
state. This is known to be true for the previously known rational and trigonometric models, and it
would therefore be very natural to verify if it is also the case for the more general elliptic
potential introduced in Section~\ref{sec.examples} or its hyperbolic limit. Another possible line
for future research is the construction and analysis of the spin versions of the near-neighbors
models considered (see Refs.~\cite{EFGR05b,EFGR07} for the rational and trigonometric models), and
their associated short-range spin chains (as was done in Ref.~\cite{EFGR08} for the rational
model). Similarly, it would be of interest to study the extension of our results to more general
Jastrow-like ground states depending on differences $x_i-x_{i+k}$ with $k$ less than a fixed range
$r>1$ (see, e.g.,~Refs.~\cite{TJK16,PBOC17}), as well as to ground states factorized over other
root systems like $BC_N$~\cite{AJK01,EGKP05}. Finally, another topic worth investigating is the
explicit computation of the correlation functions of the eigenvalue probability densities given by
the Jastrow-like ground states considered in this paper, like, e.g., the elliptic wave function in
Eq.~\eqref{psiell}. This can be done in principle with the techniques of Refs.~\cite{Gu50,Ho50},
although the evaluation of the resulting integrals could be far from trivial in this case. In
fact, the analogous problem for the density~\eqref{psifinal} with $\om=0$
and~$\chi(x)=|x|^{\be/2}$ has already been solved in Ref.~\cite{BGS99}. The corresponding
distribution of the spacings of consecutive eigenvalues has been shown in the latter reference to
be a good approximation to this statistic for certain pseudo-integrable billiards and for the
Anderson model at the transition point.

\section*{Acknowledgments}

This work was partially supported by Spain's MINECO under research grant no.\ FIS2015-63966-P. JAC
acknowledges the financial support of the Universidad Complutense de Madrid through a 2015
predoctoral scholarship. MB was supported by a grant from the Ministry of Science, Research and
Technology of Iran. She would also like to thank the Departamento de Física Teórica II of the
Universidad Complutense de Madrid for their warm hospitality.

%\bibliography{cmprefs}
%\bibliographystyle{unsrt}
%\bibliographystyle{model1a-num-names}

\end{document}